%% file: draft.tex
\documentclass[final,5p,times,twocolumn]{elsarticle}
\usepackage{lineno}
\usepackage{graphicx,color}
\usepackage{amssymb}
\usepackage{xspace}
\usepackage{amsmath}
\usepackage{latexsym}
\usepackage{amsxtra}
\usepackage{amscd}
\usepackage{dcolumn}
\usepackage{bm}
\usepackage{threeparttable}
\usepackage{dcolumn}
\usepackage{xfrac}
\usepackage{stackengine}
\newcolumntype{d}[1]{D{.}{.}{#1}}
\newcolumntype{k}{D{(}{(}{-3}}
\biboptions{numbers,sort&compress}

\usepackage[colorlinks=true, breaklinks]{hyperref}
\usepackage{url}
\usepackage{breakurl}
\graphicspath{./}

\newcommand{\nuc}[2]{$^{#1}$#2}
\newcommand{\gcm}{g/cm$^{2}$\xspace}

\newcommand{\grays}{$\gamma$ rays\xspace}
\newcommand{\ghray}{$\gamma$-ray\xspace}

\newcommand{\tpos}{$2_1^{+}$ state\xspace}

\renewcommand{\ge}{$^{62}$Ge\xspace}
\newcommand{\ga}{$^{62}$Ga\xspace}
\newcommand{\zn}{$^{62}$Zn\xspace}

\definecolor{nicegreen}{RGB}{46,204,64}

\definecolor{thblue}{RGB}{46,64,204}

\newcommand{\bev}{$B(E2;\,0_1^+ \rightarrow 2_1^+)$ value\xspace}
\newcommand{\bevs}{$B(E2;\,0_1^+ \rightarrow 2_1^+)$ values\xspace}

\journal{Physics Letters B}

\begin{document}

\begin{frontmatter}
\title{Precision Tests of Isospin Symmetry through Coulomb excitation of $A=62$ Nuclei}
\input{authors.tex}

\begin{abstract}
Isospin symmetry in the $A=62$ mass system was investigated through Coulomb excitation reactions at the RIKEN Radioactive Isotope Beam Factory. Beams of \nuc{62}{Zn}, \nuc{62}{Ga}, and \nuc{62}{Ge} were studied using the BigRIPS-ZeroDegree-DALI2$^+$ setup under identical experimental conditions, allowing for cancellation of systematic uncertainties. Inelastic scattering cross sections measured with two different targets were used to extract nuclear deformation lengths and $E2$ matrix elements. The isospin symmetry of the $A=62$ system was rigorously tested by examining the linearity of the proton matrix elements within the triplet with high precision. The observed linear relationship between the reduced proton matrix elements for the three nuclei holds within experimental uncertainties, providing a stringent test of isospin symmetry. This experiment provides the most accurate test, to date, of isospin symmetry rules using transition matrix elements. These results were interpreted using large-scale shell-model calculations, offering valuable insights into isospin symmetry behavior in this region of the nuclear chart.
\end{abstract}
\date{\today}
\end{frontmatter}
The atomic nucleus serves as a unique quantum laboratory to study many-body interactions among nucleons, protons and neutrons, mediated by the strong nuclear force. Since protons and neutrons are nearly identical apart from their electric charge and isospin quantum numbers ($t_z = \mp 1/2$, respectively), the strong interaction is assumed to be symmetric under isospin rotation, treating protons and neutrons equivalently. Consequently, states in members of isobaric multiplets with the same total isospin $T$, but differing projections $T_z = (N-Z)/2$, are expected to exhibit identical intrinsic properties. Deviations from this symmetry reveal critical insights into nuclear structure and nucleon-nucleon interactions.

Isospin symmetry is often probed through Coulomb Displacement Energies (CDEs), derived from binding energy differences within an isobaric multiplet, encapsulated in the Isobaric Multiplet Mass Equation (IMME) first introduced by Wigner~\cite{wigner57}. While the Coulomb interaction constitutes the largest contribution to CDEs, additional isospin-breaking effects from the nuclear interaction must be included to reproduce the energy levels of isobaric analog states in mirror nuclei~\cite{zuker02, bentley07}. Observables such as mirror energy differences (MED) test charge symmetry, whereas triplet energy differences (TED) probe charge independence, offering complementary insights into isospin-breaking mechanisms.

Recent experimental advances have significantly expanded spectroscopy of $J^\pi=2^+$ states in $T=1$ and $T=2$ isobaric multiplets across the nuclear chart~\cite{wimmer23, fernandez21, yajzey21}, highlighting the sensitivity of MED to nuclear structure effects~\cite{bentley07}. However, complementary probes, such as electromagnetic transition rates, are essential for a more direct evaluation of wave function components. The reduced transition probability, or $B(E2)$ value, between low-lying states provides a critical window into proton and neutron matrix elements, $M_{p}$ and $M_{n}$, respectively~\cite{bernstein79}. For the $T=1$ triplet, the \bev is directly proportional to $|M_{p}|^2$,
\begin{equation}
    B(E2;\;J_{\text{i}} \rightarrow J_{\text{f}}) = \frac{|M_{p}(J_{\text{i}} \rightarrow J_{\text{f}}) |^2}{2J_{\text{i}}+1}.
  \end{equation}
Assuming isospin conservation, the proton matrix element $M_{p}$ should scale linearly with $T_z$:  
\begin{equation}
    M_{p}(T_z) = \frac{1}{2}\left(M_0 - M_1 T_z \right)
    \label{eq:linear}
\end{equation}
where $M_0$ and $M_1$ represent the isoscalar and isovector components, respectively. The linearity of $M_p(T_z)$  is a robust test of isospin symmetry, independent of model assumptions. Deviations from linearity immediately signal a breakdown of isospin conservation in the wave functions. This relationship has been validated in light nuclei~\cite{prados07,boso19} and extended to heavier systems such as $A=70$~\cite{wimmer21}.
Notably, while linearity holds for lighter nuclei, significant deviations observed in $A=70$ were attributed to shape changes across the triplet, offering a compelling avenue for exploring the interplay between isospin symmetry and nuclear deformation. Around $A=78$ along the $N=Z$ line, strong deformation, shape coexistence, and shape changes have been previously predicted~\cite{moller09}. It is thus conceivable that small changes in the location and occupation of single-particle orbitals due to broken isospin symmetry could lead to a different ground-state shape and configuration, potentially explaining the observed increase in $M_{p}$ for \nuc{70}{Kr}~\cite{wimmer21}.

To explore the transition from the spherical nuclei around the doubly-magic \nuc{56}{Ni} toward the pronounced ground-state deformation and maximum collectivity at \nuc{76}{Sr}~\cite{lemasson12,llewellyn20}, this Letter investigates the linearity of the proton $E2$ matrix elements in the $A=62;\;T=1$ triplet nuclei \nuc{62}{Ge}, \nuc{62}{Ga}, and \nuc{62}{Zn}. Unlike previous studies of lighter nuclei, where the three members of a triplet were examined using different experimental techniques, introducing systematic uncertainties~\cite{prados07}, the present study employs identical experimental conditions for all three triplet members. This approach effectively cancels nearly all systematic uncertainties and those related to reaction modeling in relative comparisons. By leveraging this novel and unified methodology, the study achieves the highest precision test of the linearity of the proton matrix elements $M_{p}$, enabling stringent conclusions about isospin symmetry.

The experiment was conducted at the Radioactive Isotope Beam Factory, operated by the RIKEN Nishina Center and CNS, The University of Tokyo, using the BigRIPS and ZeroDegree spectrometers~\cite{kubo12} and the DALI$2^+$ array~\cite{takeuchi14}. A stable \nuc{78}{Kr} primary beam with an intensity of 300~pnA was accelerated to 345~$A$MeV and directed onto a 7-mm-thick Be target, inducing fragmentation reactions. The resulting cocktail of fragments was purified and analyzed through the BigRIPS device.
In the first stage of BigRIPS, isotopes were separated using the $B{\rho}\text{-}{\Delta}E\text{-}B{\rho}$ method. The second stage employed the ${\Delta}E \text{-} B{\rho} \text{-} TOF$ technique for precise identification of the atomic number $Z$ and mass-to-charge ratio ($A/Q$) by measuring the ions' energy loss ($\Delta E$) in an ionization chamber, trajectories using parallel plate avalanche counters, and time-of-flight (TOF) using plastic scintillators~\cite{fukuda13}.
Two BigRIPS settings were used during the experiment. The first setting focused on \nuc{62}{Ge} production, yielding 290 particles per second, and also transmitted \nuc{62}{Ga} ions at a rate of 1800~pps. The second setting was optimized for \nuc{62}{Zn}, achieving a transmission rate of 5700~pps. The purified secondary beams were then directed onto 0.48-\gcm-thick \nuc{197}{Au} and 0.26-\gcm-thick \nuc{12}{C} targets, inducing inelastic scattering reactions. The beam energies at the center of the target were around 150~$A$MeV.
Downstream of the secondary target, the ZeroDegree spectrometer was used for ion identification via the ${\Delta}E \text{-} B{\rho} \text{-} TOF$ method. Scattering angles were determined by tracking the ion trajectories before and after the target with PPACs. This setup enabled precise characterization of the reaction products and their kinematics.

\begin{figure}[!h]
  \includegraphics[width=\columnwidth]{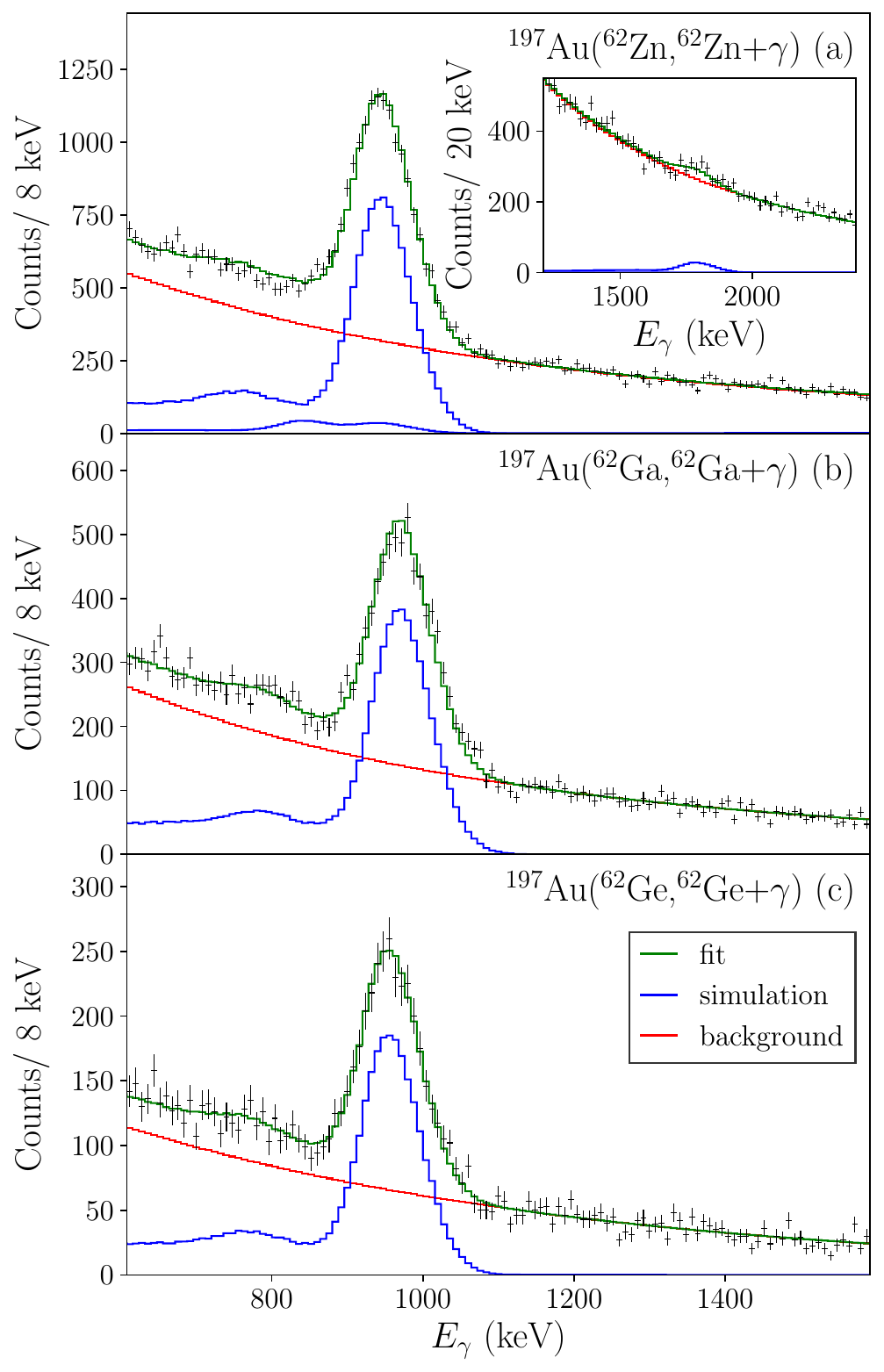}
  \caption{Doppler-corrected \ghray energy spectra for inelastic scattering on the \nuc{197}{Au} target for (a) \nuc{62}{Zn}, (b) \nuc{62}{Ga}, and (c) \nuc{62}{Ge}. The spectra were constructed using the forward-most DALI$2^+$ detectors ($\theta_\gamma < 60^{\circ}$) for background suppression. Add-back of neighboring crystals is applied. The inset shows the high-energy region of the \nuc{62}{Zn} spectrum around the $2^+_2\rightarrow 0^+_1$ transition.}
  \label{fig:gamma-ray}
\end{figure}

The emitted \grays from the ions were detected using the DALI$2^+$ array~\cite{takeuchi14}. With its high granularity of 226 NaI(Tl) detectors, the \ghray emission angles were extracted to perform Doppler correction along with the measured ion velocity. Energy and efficiency calibrations of the DALI$2^+$ detectors were carried out using standard \ghray sources.
After selecting (in)elastic scattering events by gating on the ions of interest in the BigRIPS and ZeroDegree spectrometers, Doppler correction and add-back of hits in neighboring crystals were applied. To suppress background contributions from atomic processes, only the most forward detectors, $\theta_\gamma < 60^{\circ}$, were used in the following analysis. The final spectra for the gold target are shown in Fig.~\ref{fig:gamma-ray}.
The spectra for the \nuc{12}{C} target measurements can be found in Ref.~\cite{wimmer23}. The \ghray yields were determined by fitting the measured spectra with simulated response functions generated using the Geant4 simulation toolkit~\cite{agostinelli03}, as shown in Fig.~\ref{fig:gamma-ray}. Simulations incorporated the angular distributions of $\gamma$ rays and the lifetimes of the $2^+$ states derived from the \bevs determined in this work. Additionally, a continuous double-exponential function was employed to model the beam-induced background. 
In the case of \nuc{62}{Zn}, the 1805-keV $2^+_2 \rightarrow 0^+_1$ transition was observed (see inset of Fig.~\ref{fig:gamma-ray}\,(a)). The response function for the $2^+_2$ state used in the fit included the known branching ratio of the 851~keV $2^+_2\rightarrow 2^+_1$ transition and the one to the ground state~\cite{nichols12}. For \nuc{62}{Ge} and \nuc{62}{Ga}, decays from states besides the $2^+_1$ were not observed and upper limits for indirect feeding were estimated using simulations and included in the uncertainties. 
To determine cross sections, the \ghray yields were normalized to the number of incident particles, corrected for transmission and acceptance losses in the ZeroDegree spectrometer. These corrections were higher for \nuc{62}{Ga} than for the centered beams of \nuc{62}{Ge} and \nuc{62}{Zn}. Details of the procedures and associated uncertainties are provided in Ref.~\cite{wimmer20}. The resulting cross sections for inelastic excitation of the \tpos are listed in Table~\ref{tab:results}.
\begin{table}[h]
  \caption{Summary of results for the $0^+_1 \rightarrow 2^+_1$ excitation of the studied nuclei. From the measured cross sections, deformation lengths $\delta_N$ and $M_p$ matrix elements were extracted using a reaction model calculation. Uncertainties listed for $\delta_N$ and $M_p$ are shown as statistical, systematic, and theoretical, respectively. See text for details.} 
  \label{tab:results}
  \centering
  \begin{tabular}{lccc}
    \hline
    & \nuc{62}{Ge} & \nuc{62}{Ga} & \nuc{62}{Zn}\\
    \hline
    \hline
    $E(2^+_1)$ (keV)            & 965~\cite{wimmer23}          & 977~\cite{nichols12}          & 954~\cite{nichols12} \\
    $\sigma(2^+_1)_\text{C}$ (mb)& 22.8(18) & 22.4(14)  & 22.6(13) \\
    $\Delta^\text{stat}\sigma(2^+_1)_\text{C}$ (mb)& 1.2 & 0.5  & 0.6 \\
    $\Delta^\text{syst}\sigma(2^+_1)_\text{C}$ (mb)& 1.3 & 1.3  & 1.2 \\
    $\sigma(2^+_1)_\text{Au}$ (mb)& 200(15)  & 194(25)   & 195(13) \\
    $\Delta^\text{stat}\sigma(2^+_1)_\text{Au}$ (mb)& 6  &  5  & 3 \\
    $\Delta^\text{syst}\sigma(2^+_1)_\text{Au}$ (mb)& 14  & 24   & 13 \\
    \hline
    $\delta_N$ (fm)           & 1.14(5)   &1.11(3)  & 1.13(4) \\
    $\Delta^\text{stat}\delta_N$ (fm) &0.03    &0.01   &0.02 \\
    $\Delta^\text{syst}\delta_N$ (fm) &0.03    &0.03   &0.03 \\
    \hline

    $B(E2)$ ($e^2$fm$^4$)        & 1456(184)& 1406(233)& 1405(174) \\
    $M_p$ ($e$fm$^2$)            & 38.1(24) & 37.5(31) & 37.4(23) \\
    $\Delta^\text{stat}M_p$ ($e$fm$^2$) &0.6    &0.4   &0.3  \\
    $\Delta^\text{syst}M_p$ ($e$fm$^2$) &1.2    &2.4   &1.1 \\
    $\Delta^\text{theo}M_p$ ($e$fm$^2$) &2.0    &2.0   &2.0 \\
    \hline
  \end{tabular}
  \end{table}
  Systematic uncertainties include contributions from \ghray detection efficiency (5\%), ZeroDegree acceptance and efficiency (2\% for \nuc{62}{Zn} and \nuc{62}{Ge}, 10\% for \nuc{62}{Ga}), and unobserved feeding ($<1$\%).

\begin{figure}
  \centering
  \includegraphics[width=1.0\columnwidth]{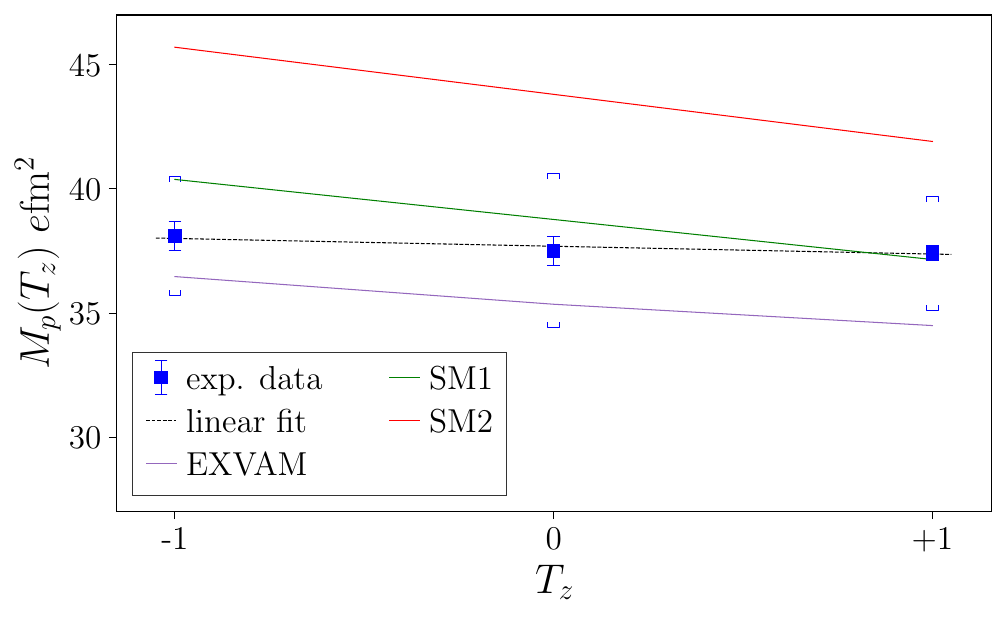}
  \caption{$M_p(T_z)$ linearity in the $A=62$ isospin triplet.  The error bars indicate statistical uncertainties, while the additional caps show the total uncertainties including statistical, systematical, and uncertainties arising from the reaction theory calculations. Large-scale shell-model calculations were performed in the $fp$ model space using the KB3GR interaction with two sets of effective charges: $e_p = 1.31$, $e_n = 0.46$ (SM1) and $e_p = 1.50$, $e_n = 0.50$ (SM2). The beyond-mean-field EXVAM calculations were done with effective charges $e_p = 1.50$, $e_n = 0.50$~\cite{mihai22}.}
  \label{fig:mptz}
\end{figure}

 The nuclear deformation lengths, ${\delta}_N$, and the $E2$ matrix elements, $M_p$, were extracted from the measured cross sections using the methodology outlined in Refs.~\cite{wimmer20}. For given values of ${\delta}_N$ and $M_p$, excitation cross sections for the reactions on C and Au targets were calculated using a modified version of the distorted wave coupled channels code Fresco~\cite{thompson88,moro18}. Optical-model potentials were derived following the procedure described in Ref.~\cite{furumoto12}, with the Coulomb potential radius parameter set to $r_C = 1.25$~fm.
 Both ${\delta}_N$ and $M_p$ were iteratively varied to reproduce the experimental cross sections for the two targets simultaneously, ensuring that both excitation modes and their interference were appropriately accounted for. The final results are presented in Table~\ref{tab:results}. 
The analysis introduces theoretical uncertainties due to the reaction modeling, specifically the neglect of dynamical relativistic corrections (5\%) and uncertainties in the optical-model potential (8\%). These theoretical  uncertainties are combined with statistical and systematic uncertainties, as listed in Table~\ref{tab:results}, with total uncertainties calculated by adding these contributions in quadrature.
The extracted proton $M_p$ matrix elements are plotted as a function of the isospin projection quantum number $T_z$ in Fig.~\ref{fig:mptz}. 
Within the experimental uncertainty, the three values of $M_p$ are nearly identical, with their most probable values lying on a straight line, reinforcing the linearity in the $A=62$ isospin triplet.

The nucleus \zn has been extensively studied, with the lifetimes of the $2^+_{1,2}$ states previously measured~\cite{ward81,kenn02,starosta07}. The adopted value for the transition matrix element is $B(E2;\;0^+_1 \rightarrow 2^+_1) = 1224(59)$~$e^2$fm$^4$ (corresponding to $M_p = 35.0(8)$~$e$fm$^2$)~\cite{pritychenko12}, which agrees well with the present results within the uncertainties.
It is worth noting that these lifetime measurements neglected feeding from states other than the $4^+_1$ state, which may have led to a slight underestimation of $M_p$. In contrast, the present work explicitly accounts for feeding contributions, resulting in a slightly higher and more reliable $M_p$ value.

The proton matrix elements can be decomposed into the isoscalar $M_0$ and isovector $M_1$ matrix elements (see Eq.~\ref{eq:linear}). A compilation of the experimentally known values is shown in Ref.~\cite{morse18}. There, they are calculated from only the $T_z=0,1$ members. The present data for \zn and \ga yields $M_0 = 75.0(12)$~$e$fm$^2$ and $M_1 = -0.2(13)$~$e$fm$^2$. If the newly measured value for \ge is included, the matrix elements amount to $M_0 = 75.4(4)$~$e$fm$^2$ and $M_1 = +0.6(4)$~$e$fm$^2$ fully in line with the systematics.

To interpret the experimental data, large-scale shell-model calculations were performed using the KB3GR effective interaction~\cite{kb3gr} combined with the Coulomb interaction in the $pf$ model space. The calculations utilized the ANTOINE code~\cite{caurier05}, allowing for up to $t = 8$ nucleons to be excited from the $f_{7/2}$ orbital to the upper orbits. Two sets of quadrupole effective charges were employed: $e_{\pi} = 1.31e$, $e_{\nu} = 0.46e$ (SM1) and $e_{\pi} = 1.5e$, $e_{\nu} = 0.5e$ (SM2). The first set of charges was microscopically derived for harmonic oscillator cores~\cite{dufour96,lenzi21}, while the second represents the standard values commonly used in shell-model studies in this model space.
The results for the proton $M_p$ matrix elements are displayed in Fig.~\ref{fig:mptz}. It can be seen that the microscopically derived effective charges better reproduce the experimental data. Additionally, theoretical results from Ref.~\cite{mihai22} are shown in Fig.~\ref{fig:mptz} for comparison. These are beyond-mean-field complex excited Vampir (EXVAM) calculations, which use a \nuc{40}{Ca} core and include the oscillator orbits up to $0g_{9/2}$ for both protons and neutrons. The EXVAM calculations predict a linear trend in the proton $M_p$ matrix elements and successfully reproduce their overall magnitude.

\begin{figure}
    \centering
    \includegraphics[width=\columnwidth]{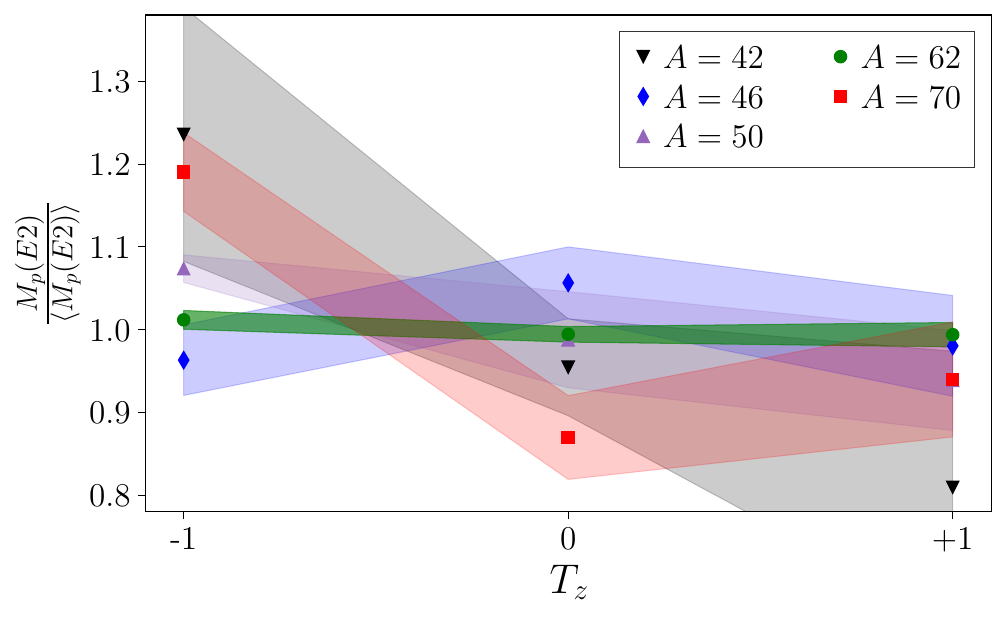}
    \caption{Linearity of $M_p(T_z) / \langle M_p \rangle$ in $A=42 - 70$ mass triplets. The results of the present study are compared to values from the literature~\cite{hadynska16,boso19,giles19,wimmer21,pritychenko16}. The bands indicate the corresponding uncertainties. The linearity test is performed relatively within each mass triplet. For $A=62$, the error band represents statistical uncertainties including the small transmission correction. For all other triplets, total uncertainties (statistical and systematic) are used in the comparison.}

    \label{fig:mptzaverage}
  \end{figure}

While the matrix elements for the analogous $2^+_1 \rightarrow 0^+_1$ transitions of the isobaric triplet members follow the linear trend predicted by isospin symmetry~\cite{bernstein79}, the present experimental results allow for a more stringent test of this symmetry. Since the data were acquired under identical experimental conditions for all three nuclei, nearly all systematic uncertainties cancel in a relative comparison of the proton matrix elements. This unique approach enables a high-precision evaluation.
To account for the variations in absolute values of $M_p$ along the $N=Z$ line, the deviation from the average value $\langle M_p \rangle$ is plotted in Fig.~\ref{fig:mptzaverage}.
To rigorously test the linearity, only statistical uncertainties and a minor contribution from the transmission correction for \ga need to be considered. The uncertainty band for the $A=62;\;T=1$ triplet permits a highly stringent test of isospin symmetry at the percent level. The average uncertainty across the triplet is 1.2\%, representing at least a fourfold improvement in precision over lower mass $T=1$ triplets. 
For comparison, Fig.~\ref{fig:mptzaverage} also displays results from other $A=42 - 70$ isospin triplets~\cite{hadynska16,boso19,giles19,wimmer21,pritychenko16}.

\begin{figure}
    \centering
    \includegraphics[width=\columnwidth]{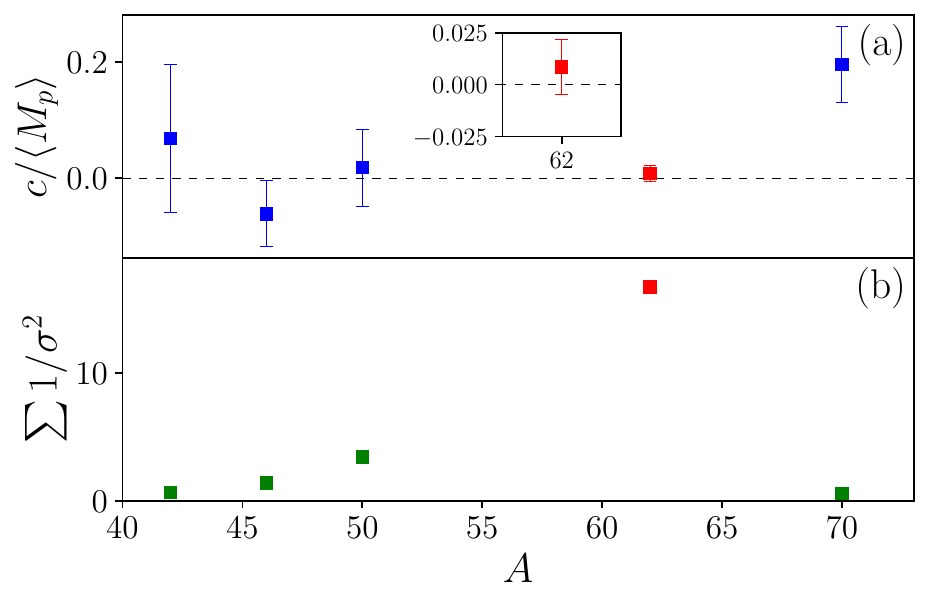}
    \caption{Metric quantifying the linearity of $M_p$ and the data quality. (a) $c$ coefficient from a quadratic fit $M_p=a+b\cdot T_z+c\cdot T_z^2$ for $T=1$ triplets. (b) Information content for the same triplets. The results from the present study are highlighted in red.}
    \label{fig:score}
\end{figure}

To test further the linearity rule for transition matrix elements in isospin multiplets, one can fit a quadratic function of the form $M_p=a+b\cdot T_z+c\cdot T_z^2$ and evaluate the coefficient $c$, which should vanish if isospin symmetry holds exactly. This approach has already been used for isospin triplets up to $A=46$~\cite{boso19}. Fig.~\ref{fig:score}\,(a) shows the value of $c$ normalized to the average matrix elements for easier comparison. 
For the present case, the value of $c$ is fully consistent with zero, and the new data provides the most stringent test of the linearity of proton matrix elements to date.  
The ability to validate isospin symmetry can also be quantified using the information content, defined as $\sum_i 1 / \sigma_i^2$. This metric is shown in Fig.~\ref{fig:score}\,(b) and compared to other $T=1$ triplets. The comparison confirms that the quality of the present data surpasses all previous tests of linearity.


The results of this study should be compared to the $3\sigma$ deviation from the linear, isospin-conserving $M_p$ relation observed for $A=70$~\cite{wimmer21}. This deviation is associated with \nuc{70}{Kr}, which lies at the onset of strong ground-state deformation maximized in the Sr and Zr isotopes~\cite{lemasson12,llewellyn20}. Additionally, \nuc{70}{Kr} is situated in a region where shape coexistence and dramatic shape changes are known to occur~\cite{bouchez03}. 
In contrast, the nuclei investigated in this work are only weakly collective and are well described by the shell model, as illustrated in Fig.~\ref{fig:mptz}. 
Future experiments focusing on the intermediate $A=66$ triplet and the presumably much more collective $A=74$ or heavier systems could further probe this interpretation. Such studies would provide crucial insights into the interplay between isospin symmetry and deformation effects in nuclear structure.

In summary, we have performed relativistic Coulomb and nuclear inelastic scattering experiments on the $A=62$ isospin triplet nuclei \nuc{62}{Zn}, \nuc{62}{Ga}, and \nuc{62}{Ge} using both light and heavy nuclear targets. The measured cross sections were analyzed to consistently extract nuclear deformation lengths and $E2$ matrix elements. This approach enabled the most accurate test of isospin symmetry rules using transition matrix elements to date. The deduced proton matrix elements $M_p$ exhibit a linear dependence on the isospin projection $T_z$, in excellent agreement with large-scale shell-model calculations. By utilizing nearly identical experimental conditions for all three nuclei, isospin symmetry could be tested with an unprecedented precision at the percent level, which is a remarkable achievement for experiments with low-intensity, rare isotope beams. Our findings confirm the preservation of isospin symmetry in weakly collective nuclei, reinforcing the consistency of nuclear wave functions across the triplet. These results provide a stringent benchmark for theory and suggest that deviations observed in heavier, more collective nuclei are driven by deformation effects, which amplify small symmetry-breaking contributions. This study underscores the importance of systematic and high-precision investigations to disentangle the complex interplay between isospin symmetry and nuclear structure.

We would like to thank the RIKEN accelerator and the BigRIPS teams for providing the high intensity beams. 
K.~W. acknowledges the support from the Spanish Ministerio de Ciencia, Innovaci\'on y Universidades grant RYC-2017-22007 and the European Research Council through the ERC Grant No. 101001561-LISA. A.~P. is supported in part by grants CEX2020-001007-S  funded by  MCIN/AEI (Spain) /10.13039/501100011033 and PID2021-127890NB-I00. F.~B. was supported by the RIKEN Special Postdoctoral Researcher Program. A.~J. acknowledges funding through project PID2023-150056NB-C42 financed by MICIU/AEI /10.13039/501100011033 and by FEDER, UE. T.~F. is supported by Japan Society for the Promotion of Science (JSPS) KAKENHI Grant Numbers JP20K03944. B.~M. was an International Research Fellow of the Japanese Society for the Promotion of Science. The work is further supported by the UK STFC under Grants Nos. ST/L005727/1, ST/P003885/1, ST/V001035/1.
\bibliographystyle{elsarticle-num-names}
\bibliography{draft}

\end{document}

%% file: authors.tex
\author[gsi,ut,rnc]{K.~Wimmer}
\cortext[cor1]{Corresponding author}
\ead{k.wimmer@gsi.de}
\author[ific]{T.~H{\"u}y{\"u}k}
\author[upa,infn]{S.~M.~Lenzi}
\author[uam]{A.~Poves}
\author[rnc]{F.~Browne}
\author[rnc]{P.~Doornenbal}
\author[ut,rnc]{T.~Koiwai}
\author[gsi]{T.~Arici}
\author[york]{M.~A.~Bentley}
\author[infn]{M.~L.~Cort\'es}
\author[ynu]{T.~Furumoto}
\author[cns]{N.~Imai}
\author[csic]{A.~Jungclaus}
\author[cns]{N.~Kitamura}
\author[msu]{B.~Longfellow}
\author[cnrs]{R.~Lozeva}
\author[rnc]{B.~Mauss}
\author[infn]{D.~Napoli}
\author[ut]{M.~Niikura}
\author[york]{X.~Pereira-Lopez}
\author[upa,infn]{F.~Recchia}
\author[jyfl]{P.~Ruotsalainen}
\author[ut,rnc]{R.~Taniuchi}
\author[york]{S.~Uthayakumaar}
\author[csic]{V.~Vaquero}
\author[york]{R.~Wadsworth}
\author[jazan]{R.~Yajzey}

\address[gsi]{GSI Helmholtzzentrum f\"{u}r Schwerionenforschung, D-64291 Darmstadt, Germany}
\address[ut]{Department of Physics, The University of Tokyo, Hongo, Bunkyo-ku, Tokyo 113-0033, Japan}
\address[rnc]{RIKEN Nishina Center, 2-1 Hirosawa, Wako, Saitama 351-0198, Japan}
\address[ific]{Instituto de Fisica Corpuscular, CSIC-Universidad de Valencia, E-46071 Valencia, Spain}
\address[upa]{Dipartimento di Fisica e Astronomia, Universit\`a di Padova, Padova I-35131, Italy}  
\address[infn]{INFN, Sezione di Padova, Padova I-35131, Italy} 
\address[uam]{Departamento de F\'isica Te\'orica, Universidad Aut\'onoma de Madrid, 28049 Madrid, Spain}
\address[york]{School of Physics, Engineering and Technology, University of York, YO10 5DD York, United Kingdom}
\address[ynu]{College of Education, Yokohama National University, Yokohama 240-8501, Japan}
\address[cns]{Center for Nuclear Study, University of Tokyo, Hongo, Bunkyo-ku, Tokyo 113-0033, Japan}
\address[csic]{Instituto de Estructura de la Materia, CSIC, E-28006 Madrid, Spain}
\address[msu]{National Superconducting Cyclotron Laboratory and Department of Physics and Astronomy, Michigan State University, East Lansing, MI 48824 USA}
\address[cnrs]{Université Paris-Saclay, CNRS\slash IN2P3, IJCLab, 91405 Orsay, France}
\address[jyfl]{Accelerator Laboratory, Department of Physics, University of Jyv\"askyl\"a, FI-40014 Jyv\"askyl\"a, Finland}
\address[jazan]{Department of Physical Sciences, Physics Division, College of Science, Jazan University, P.O. Box. 114, Jazan 45142, Kingdom of Saudi Arabia}